\RequirePackage{lineno}
\documentclass[aps,prl,showpacs,letterpaper,twocolumn,longbibliography,superscriptaddress]{revtex4-1}%
\usepackage{graphicx}
\usepackage{bm}
\usepackage{epsf}
\usepackage{rotating}
\usepackage{epsfig,graphics,rotate,color}
\usepackage{wrapfig}
\usepackage{amssymb}
\usepackage{amsmath}
\usepackage{amsfonts}
\usepackage{array,hhline,dcolumn}%
\usepackage{gensymb}
\usepackage{placeins}
\usepackage{tabularx}
\usepackage[colorlinks=true,citecolor=blue,linkcolor=blue]{hyperref}
\usepackage{multirow, makecell}
\usepackage{booktabs}
\usepackage{siunitx}
\usepackage{adjustbox}

\usepackage[utf8]{inputenc}

\usepackage{accents}
\usepackage{color}
\usepackage{xcolor}
\usepackage[normalem]{ulem}

\usepackage{txfonts}
\DeclareMathSymbol{\diamonddot}{\mathord}{symbolsC}{144}

\def\al{\alpha}
\def\be{\beta}
\def\ga{\gamma}
\def\de{\delta}

\def\th{\theta}

\def\ta{\tau}

\def\ph{\phi}

\def\De{\Delta}

\def\nue{\nu_e}
\def\numu{\nu_\mu}
\def\nutau{\nu_\tau}

\def\nuebar{\bar\nu_e}
\def\numubar{\bar\nu_\mu}
\def\nutaubar{\bar\nu_\tau}

\def\ket#1{|{#1}\rangle}

\def\lsim{\mathrel{\rlap{\lower4pt\hbox{\hskip1pt$\sim$}}
    \raise1pt\hbox{$<$}}}
\def\gsim{\mathrel{\rlap{\lower4pt\hbox{\hskip1pt$\sim$}}
    \raise1pt\hbox{$>$}}}
\def\Re{\hbox{Re}\,}
\def\Im{\hbox{Im}\,}

\newcommand{\beq}{\begin{eqnarray}}
\newcommand{\eeq}{\end{eqnarray}}
\def\to{\rightarrow}

\newcommand*{\Scale}[2][4]{\scalebox{#1}{\ensuremath{#2}}}%

\newcommand{\GeV}{\si{\giga\electronvolt}}

\def\adiii{\accentset{\circ}{a}^{(3)}} 
\def\cdiv{\accentset{\circ}{c}^{(4)}} 
\def\adv{\accentset{\circ}{a}^{(5)}}
\def\cdvi{\accentset{\circ}{c}^{(6)}}
\def\advii{\accentset{\circ}{a}^{(7)}}
\def\cdviii{\accentset{\circ}{c}^{(8)}}

\def\aeediii{\accentset{\circ}{a}^{(3)}_{e e}} 
\def\ceediv{\accentset{\circ}{c}^{(4)}_{e e}} 
\def\aeedv{\accentset{\circ}{a}^{(5)}_{e e}} 
\def\ceedvi{\accentset{\circ}{c}^{(6)}_{e e}} 
\def\aeedvii{\accentset{\circ}{a}^{(7)}_{e e}} 
\def\ceedviii{\accentset{\circ}{c}^{(8)}_{e e}}

\def\aeudiii{\accentset{\circ}{a}^{(3)}_{e \mu}} 
\def\ceudiv{\accentset{\circ}{c}^{(4)}_{e \mu}} 
\def\aeudv{\accentset{\circ}{a}^{(5)}_{e \mu}} 
\def\ceudvi{\accentset{\circ}{c}^{(6)}_{e \mu}} 
\def\aeudvii{\accentset{\circ}{a}^{(7)}_{e \mu}} 
\def\ceudviii{\accentset{\circ}{c}^{(8)}_{e \mu}}

\def\aetdiii{\accentset{\circ}{a}^{(3)}_{e \tau}} 
\def\cetdiv{\accentset{\circ}{c}^{(4)}_{e \tau}} 
\def\aetdv{\accentset{\circ}{a}^{(5)}_{e \tau}} 
\def\cetdvi{\accentset{\circ}{c}^{(6)}_{e \tau}} 
\def\aetdvii{\accentset{\circ}{a}^{(7)}_{e \tau}} 
\def\cetdviii{\accentset{\circ}{c}^{(8)}_{e \tau}}

\def\auudiii{\accentset{\circ}{a}^{(3)}_{\mu\mu}} 
\def\cuudiv{\accentset{\circ}{c}^{(4)}_{\mu\mu}} 
\def\auudv{\accentset{\circ}{a}^{(5)}_{\mu\mu}}
\def\cuudvi{\accentset{\circ}{c}^{(6)}_{\mu\mu}}
\def\auudvii{\accentset{\circ}{a}^{(7)}_{\mu\mu}}
\def\cuudviii{\accentset{\circ}{c}^{(8)}_{\mu\mu}}

\def\autdiii{\accentset{\circ}{a}^{(3)}_{\mu\tau}} 
\def\cutdiv{\accentset{\circ}{c}^{(4)}_{\mu\tau}} 
\def\autdv{\accentset{\circ}{a}^{(5)}_{\mu\tau}}
\def\cutdvi{\accentset{\circ}{c}^{(6)}_{\mu\tau}}
\def\autdvii{\accentset{\circ}{a}^{(7)}_{\mu\tau}}
\def\cutdviii{\accentset{\circ}{c}^{(8)}_{\mu\tau}}

\def\attdiii{\accentset{\circ}{a}^{(3)}_{\tau\tau}} 
\def\cttdiv{\accentset{\circ}{c}^{(4)}_{\tau\tau}} 
\def\attdv{\accentset{\circ}{a}^{(5)}_{\tau\tau}}
\def\cttdvi{\accentset{\circ}{c}^{(6)}_{\tau\tau}}
\def\attdvii{\accentset{\circ}{a}^{(7)}_{\tau\tau}}
\def\cttdviii{\accentset{\circ}{c}^{(8)}_{\tau\tau}}

\def\BFSubstantial{10}
\def\BFStrong{31.6}

\def\baylimdimiiitaeufowo{6 \times 10^{-26}}%
\def\baylimdimiiitaetfowo{3 \times 10^{-27}}%
\def\baylimdimiiitauufowo{3 \times 10^{-27}}%
\def\baylimdimiiitattfowo{5 \times 10^{-27}}%

\def\baylimdimiiitaeefwoo{4 \times 10^{-28}}%
\def\baylimdimiiitautfwoo{6 \times 10^{-27}}
\def\baylimdimiiitattfwoo{2 \times 10^{-27}}

\def\baylimdimivtceufowo{2 \times 10^{-31}}
\def\baylimdimivtcetfowo{7 \times 10^{-33}}
\def\baylimdimivtcuufowo{4 \times 10^{-33}}
\def\baylimdimivtcttfowo{1 \times 10^{-32}}

\def\baylimdimivtceefwoo{6 \times 10^{-33}}%
\def\baylimdimivtcutfwoo{7 \times 10^{-34}}
\def\baylimdimivtcttfwoo{8 \times 10^{-34}}

\def\baylimdimvtaeufowo{3 \times 10^{-36}}%
\def\baylimdimvtaetfowo{9 \times 10^{-39}}%
\def\baylimdimvtauufowo{8 \times 10^{-39}}%
\def\baylimdimvtattfowo{3 \times 10^{-38}}%

\def\baylimdimvtattfwto{2 \times 10^{-35}}%

\def\baylimdimvtaeefwoo{7 \times 10^{-40}}%
\def\baylimdimvtautfwoo{4 \times 10^{-39}}
\def\baylimdimvtattfwoo{2 \times 10^{-38}}

\def\baylimdimvitceufowo{4 \times 10^{-41}}
\def\baylimdimvitcetfowo{3 \times 10^{-44}}
\def\baylimdimvitcuufowo{7 \times 10^{-45}}
\def\baylimdimvitcttfowo{1 \times 10^{-43}}

\def\baylimdimvitcttfwto{3 \times 10^{-36}}

\def\baylimdimvitceefwoo{2 \times 10^{-44}}%

\def\baylimdimvitcutfwoo{6 \times 10^{-45}}
\def\baylimdimvitcttfwoo{6 \times 10^{-45}}

\def\baylimdimviitaeufowo{5 \times 10^{-46}}
\def\baylimdimviitaetfowo{4 \times 10^{-50}}
\def\baylimdimviitauufowo{4 \times 10^{-50}}
\def\baylimdimviitattfowo{2 \times 10^{-49}}

\def\baylimdimviitattfwto{3 \times 10^{-45}}

\def\baylimdimviitaeefwoo{8 \times 10^{-51}}%
\def\baylimdimviitautfwoo{2 \times 10^{-49}}
\def\baylimdimviitattfwoo{3 \times 10^{-49}}

\def\baylimdimviiitceufowo{1 \times 10^{-50}}
\def\baylimdimviiitcetfowo{6 \times 10^{-56}}
\def\baylimdimviiitcuufowo{5 \times 10^{-56}}
\def\baylimdimviiitcttfowo{6 \times 10^{-55}}

\def\baylimdimviiitcttfwto{3 \times 10^{-49}}

\def\baylimdimviiitceefwoo{3 \times 10^{-55}}%
\def\baylimdimviiitcutfwoo{5 \times 10^{-55}}
\def\baylimdimviiitcttfwoo{8 \times 10^{-56}}

\def\NDOM{5,160}
\def\NString{86}
\def\BString{1,450}
\def\EString{2,450}
\def\DOMseparation{17}
\def\DHole{125}

\def\PMTsize{25.4}

\def\Nevt{60}
\def\Ncas{41}
\def\Ntrk{17}
\def\Ndub{{\rm two}}

\def\NbinC{10}
\def\HbinC{+1.0}
\def\LbinC{-1.0}
\def\NbinE{20}
\def\LbinE{60}
\def\HbinE{2}
\def\NbinL{10}
\def\LbinL{10}
\def\HbinL{100}
\def\conone{68}
\def\contwo{95}

\def\ThSolC{0.304}
\def\ThSolL{-0.012}
\def\ThSolH{+0.013}
\def\ThAtmC{0.570}
\def\ThAtmL{-0.024}
\def\ThAtmH{+0.018}
\def\ThReaC{0.02221}
\def\ThReaL{-0.00062}
\def\ThReaH{+0.00068}
\def\DmSolC{7.42}
\def\DmSolL{-0.20}
\def\DmSolH{+0.21}
\def\DmAtmC{2.514}
\def\DmAtmL{-0.027}
\def\DmAtmH{+0.028}

\def\SNconvEE{40}

\def\SNmuonEE{50}

\def\SDOMEE{10}

\def\SDOMangEE{50}

\def\SIceEE{20}

\def\chain{\sim 18,000}
\def\livep{800}
\def\toler{0.05}

\def\Planck{1.22\times 10^{19}}

\def\ifratio{\phi_e^i:\phi_\mu^i:\phi_\tau^i}
\def\eratio{(1\!:\!0\!:\!0)_S}
\def\muratio{(0\!:\!1\!:\!0)_S}
\def\piratio{(1/3\!:\!2/3\!:\!0)_S}

\begin{document}

\title{Search for Quantum Gravity Using Astrophysical Neutrino Flavour with IceCube}

\affiliation{III. Physikalisches Institut, RWTH Aachen University, D-52056 Aachen, Germany}
\affiliation{Department of Physics, University of Adelaide, Adelaide, 5005, Australia}
\affiliation{Dept. of Physics and Astronomy, University of Alaska Anchorage, 3211 Providence Dr., Anchorage, AK 99508, USA}
\affiliation{Dept. of Physics, University of Texas at Arlington, 502 Yates St., Science Hall Rm 108, Box 19059, Arlington, TX 76019, USA}
\affiliation{CTSPS, Clark-Atlanta University, Atlanta, GA 30314, USA}
\affiliation{School of Physics and Center for Relativistic Astrophysics, Georgia Institute of Technology, Atlanta, GA 30332, USA}
\affiliation{Dept. of Physics, Southern University, Baton Rouge, LA 70813, USA}
\affiliation{Dept. of Physics, University of California, Berkeley, CA 94720, USA}
\affiliation{Lawrence Berkeley National Laboratory, Berkeley, CA 94720, USA}
\affiliation{Institut f{\"u}r Physik, Humboldt-Universit{\"a}t zu Berlin, D-12489 Berlin, Germany}
\affiliation{Fakult{\"a}t f{\"u}r Physik {\&} Astronomie, Ruhr-Universit{\"a}t Bochum, D-44780 Bochum, Germany}
\affiliation{Universit{\'e} Libre de Bruxelles, Science Faculty CP230, B-1050 Brussels, Belgium}
\affiliation{Vrije Universiteit Brussel (VUB), Dienst ELEM, B-1050 Brussels, Belgium}
\affiliation{Department of Physics and Laboratory for Particle Physics and Cosmology, Harvard University, Cambridge, MA 02138, USA}
\affiliation{Dept. of Physics, Massachusetts Institute of Technology, Cambridge, MA 02139, USA}
\affiliation{Dept. of Physics and Institute for Global Prominent Research, Chiba University, Chiba 263-8522, Japan}
\affiliation{Department of Physics, Loyola University Chicago, Chicago, IL 60660, USA}
\affiliation{Dept. of Physics and Astronomy, University of Canterbury, Private Bag 4800, Christchurch, New Zealand}
\affiliation{Dept. of Physics, University of Maryland, College Park, MD 20742, USA}
\affiliation{Dept. of Astronomy, Ohio State University, Columbus, OH 43210, USA}
\affiliation{Dept. of Physics and Center for Cosmology and Astro-Particle Physics, Ohio State University, Columbus, OH 43210, USA}
\affiliation{Niels Bohr Institute, University of Copenhagen, DK-2100 Copenhagen, Denmark}
\affiliation{Dept. of Physics, TU Dortmund University, D-44221 Dortmund, Germany}
\affiliation{Dept. of Physics and Astronomy, Michigan State University, East Lansing, MI 48824, USA}
\affiliation{Dept. of Physics, University of Alberta, Edmonton, Alberta, Canada T6G 2E1}
\affiliation{Erlangen Centre for Astroparticle Physics, Friedrich-Alexander-Universit{\"a}t Erlangen-N{\"u}rnberg, D-91058 Erlangen, Germany}
\affiliation{Physik-department, Technische Universit{\"a}t M{\"u}nchen, D-85748 Garching, Germany}
\affiliation{D{\'e}partement de physique nucl{\'e}aire et corpusculaire, Universit{\'e} de Gen{\`e}ve, CH-1211 Gen{\`e}ve, Switzerland}
\affiliation{Dept. of Physics and Astronomy, University of Gent, B-9000 Gent, Belgium}
\affiliation{Dept. of Physics and Astronomy, University of California, Irvine, CA 92697, USA}
\affiliation{Karlsruhe Institute of Technology, Institute for Astroparticle Physics, D-76021 Karlsruhe, Germany }
\affiliation{Karlsruhe Institute of Technology, Institute of Experimental Particle Physics, D-76021 Karlsruhe, Germany }
\affiliation{Dept. of Physics, Engineering Physics, and Astronomy, Queen's University, Kingston, ON K7L 3N6, Canada}
\affiliation{Dept. of Physics and Astronomy, University of Kansas, Lawrence, KS 66045, USA}
\affiliation{Dept. of Physics, King's College London, London WC2R 2LS, UK}
\affiliation{Dept. of Physics and Astronomy, Queen Mary University of London, London E1 4NS, UK}
\affiliation{Department of Physics and Astronomy, UCLA, Los Angeles, CA 90095, USA}
\affiliation{Centre for Cosmology, Particle Physics and Phenomenology - CP3, Universit{\'e} catholique de Louvain, Louvain-la-Neuve, Belgium}
\affiliation{Department of Physics, Mercer University, Macon, GA 31207-0001, USA}
\affiliation{Dept. of Astronomy, University of Wisconsin{\textendash}Madison, Madison, WI 53706, USA}
\affiliation{Dept. of Physics and Wisconsin IceCube Particle Astrophysics Center, University of Wisconsin{\textendash}Madison, Madison, WI 53706, USA}
\affiliation{Institute of Physics, University of Mainz, Staudinger Weg 7, D-55099 Mainz, Germany}
\affiliation{Department of Physics, Marquette University, Milwaukee, WI, 53201, USA}
\affiliation{Institut f{\"u}r Kernphysik, Westf{\"a}lische Wilhelms-Universit{\"a}t M{\"u}nster, D-48149 M{\"u}nster, Germany}
\affiliation{Bartol Research Institute and Dept. of Physics and Astronomy, University of Delaware, Newark, DE 19716, USA}
\affiliation{Dept. of Physics, Yale University, New Haven, CT 06520, USA}
\affiliation{Dept. of Physics, University of Oxford, Parks Road, Oxford OX1 3PU, UK}
\affiliation{Dept. of Physics, Drexel University, 3141 Chestnut Street, Philadelphia, PA 19104, USA}
\affiliation{Physics Department, South Dakota School of Mines and Technology, Rapid City, SD 57701, USA}
\affiliation{Dept. of Physics, University of Wisconsin, River Falls, WI 54022, USA}
\affiliation{Dept. of Physics and Astronomy, University of Rochester, Rochester, NY 14627, USA}
\affiliation{Department of Physics and Astronomy, University of Utah, Salt Lake City, UT 84112, USA}
\affiliation{Oskar Klein Centre and Dept. of Physics, Stockholm University, SE-10691 Stockholm, Sweden}
\affiliation{Dept. of Physics and Astronomy, Stony Brook University, Stony Brook, NY 11794-3800, USA}
\affiliation{Dept. of Physics, Sungkyunkwan University, Suwon 16419, Korea}
\affiliation{Institute of Basic Science, Sungkyunkwan University, Suwon 16419, Korea}
\affiliation{Dept. of Physics and Astronomy, University of Alabama, Tuscaloosa, AL 35487, USA}
\affiliation{Dept. of Astronomy and Astrophysics, Pennsylvania State University, University Park, PA 16802, USA}
\affiliation{Dept. of Physics, Pennsylvania State University, University Park, PA 16802, USA}
\affiliation{Dept. of Physics and Astronomy, Uppsala University, Box 516, S-75120 Uppsala, Sweden}
\affiliation{Dept. of Physics, University of Wuppertal, D-42119 Wuppertal, Germany}
\affiliation{DESY, D-15738 Zeuthen, Germany}

\author{R. Abbasi}
\affiliation{Department of Physics, Loyola University Chicago, Chicago, IL 60660, USA}
\author{M. Ackermann}
\affiliation{DESY, D-15738 Zeuthen, Germany}
\author{J. Adams}
\affiliation{Dept. of Physics and Astronomy, University of Canterbury, Private Bag 4800, Christchurch, New Zealand}
\author{J. A. Aguilar}
\affiliation{Universit{\'e} Libre de Bruxelles, Science Faculty CP230, B-1050 Brussels, Belgium}
\author{M. Ahlers}
\affiliation{Niels Bohr Institute, University of Copenhagen, DK-2100 Copenhagen, Denmark}
\author{M. Ahrens}
\affiliation{Oskar Klein Centre and Dept. of Physics, Stockholm University, SE-10691 Stockholm, Sweden}
\author{J.M. Alameddine}
\affiliation{Dept. of Physics, TU Dortmund University, D-44221 Dortmund, Germany}
\author{C. Alispach}
\affiliation{D{\'e}partement de physique nucl{\'e}aire et corpusculaire, Universit{\'e} de Gen{\`e}ve, CH-1211 Gen{\`e}ve, Switzerland}
\author{A. A. Alves Jr.}
\affiliation{Karlsruhe Institute of Technology, Institute for Astroparticle Physics, D-76021 Karlsruhe, Germany }
\author{N. M. Amin}
\affiliation{Bartol Research Institute and Dept. of Physics and Astronomy, University of Delaware, Newark, DE 19716, USA}
% ...

\collaboration{IceCube Collaboration}
\noaffiliation


\maketitle

Along their long propagation from production to detection, neutrino states undergo quantum interference which converts their types, or flavours~\cite{Super-Kamiokande:1998kpq,SNO:2001kpb}.
High-energy astrophysical neutrinos, first observed by the IceCube Neutrino Observatory~\cite{IceCube:2013low}, are known to propagate unperturbed over a billion light years in vacuum~\cite{IceCube:2018cha}.
These neutrinos act as the largest quantum interferometer and are sensitive to the smallest effects in vacuum due to new physics.
Quantum gravity (QG)~\cite{Hawking:1982dj} aims to describe gravity in a quantum mechanical framework, unifying matter, forces and space-time. 
QG effects are expected to appear at the ultra-high-energy scale known as the Planck energy, $E_{P}\equiv\Planck$~giga-electronvolts (GeV).
Such a high-energy universe would have existed only right after the Big Bang and it is inaccessible by human technologies. 
On the other hand, it is speculated that the effects of QG may exist in our
low-energy vacuum~\cite{Kostelecky:1988zi,Amelino-Camelia:1997ieq,Pospelov:2010mp,Addazi:2021xuf}, but are suppressed by the Planck energy as $E_{P}^{-1}$ ($\sim 10^{-19}$~GeV$^{-1}$), $E_{P}^{-2}$ ($\sim 10^{-38}$~GeV$^{-2}$), or its higher powers.
Measuring the coupling of particles to such small effects is difficult via kinematic observables, but the phase shift of neutrino waves could result in a sizeable flavour conversions.
Here, we report the first result of neutrino interferometry~\cite{IceCube:2017qyp} using astrophysical neutrino flavours~\cite{IceCube:2020wum,IceCube:2020abv} to search for new space-time structure. 
We did not find any evidence of anomalous flavour conversion in IceCube astrophysical neutrino flavour data. 
We place the most stringent limits of any known technologies, down to $10^{-42}$~GeV$^{-2}$ with Bayes factor (BF) $>\BFSubstantial$, on the dimension-six operators that parameterize the space-time defects. 
For the first time, by using astrophysical neutrino flavour interferometery, we unambiguously reach the signal region of quantum-gravity motivated physics.

\begin{figure}
\centering
\includegraphics[width=0.9\columnwidth]{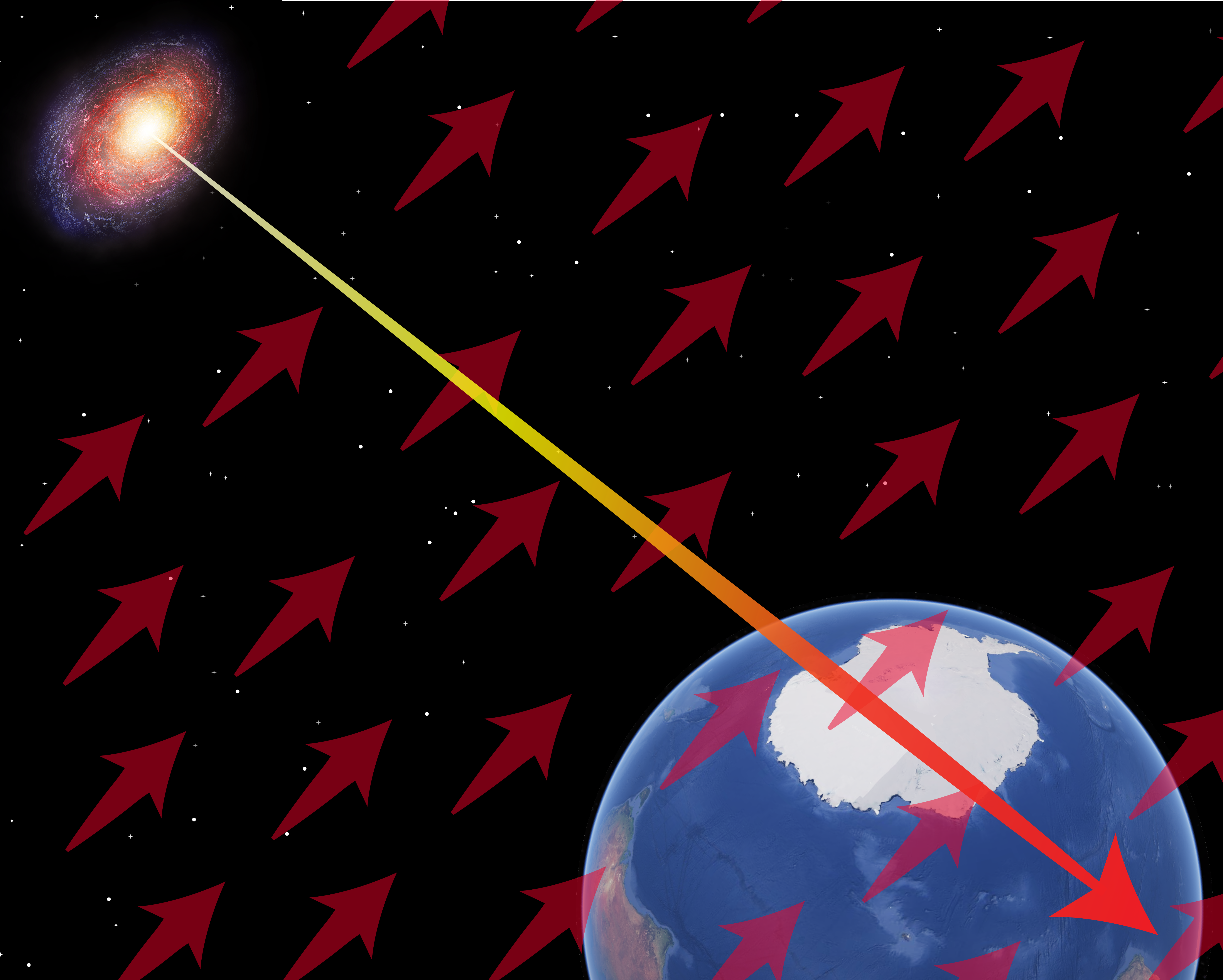}
\caption{
\textbf{\textit{Illustration of this analysis.}} 
High-energy astrophysical neutrinos, from $\LbinE$~TeV to $\HbinE$~PeV, are emitted from distant high-energy sources (shown by a long arrow).
The neutrino propagation may be affected by space-time defects which can be viewed as an  \ae ther-like media in vacuum, and in general these defects have directions depicted by red arrows. 
Although the effect may depend on directions, we assume it is isotropic in our frame without loss of generality as data is compatible with an isotropic distribution~\cite{IceCube:2020wum}.
}
\label{fig:cartoon}
\end{figure}

In the past, QG was searched for through astrophysical neutrino spectrum distortion~\cite{Diaz:2013wia} and time-of-flight~\cite{Ellis:2018ogq,Amelino-Camelia:2016ohi,Huang:2018ham} measurements. 
Here, we focus on astrophysical neutrino flavour information to search for QG effects via neutrino interferometry. 
Neutrino interferometry~\cite{IceCube:2017qyp} has been applied in various terrestrial neutrino experiments; however no evidence of QG has been found~\cite{Kostelecky:2008ts}.
The sensitivity of neutrino interferometry for astrophysical neutrinos exceeds any terrestrial experiments because of the higher energies and longer propagation distances of the neutrinos involved; see Fig.~\ref{fig:cartoon} for an illustration.

Neutrino interactions with low-energy manifestations of QG can be modelled using effective operators~\cite{IceCube:2017qyp}, such as
\beq
H\sim\frac{m^2}{2E}+\adiii-E\cdot\cdiv+E^2\cdot\adv-E^3\cdot\cdvi\cdots .
\label{eq:hamiltonian}
\eeq
The first term in this Hamiltonian describes the neutrino mass term~\cite{Esteban:2020cvm}, where we  assume the normal mass ordering as both mass ordering assumption provide comparable results. 
All other terms ($\adiii$, $\cdiv$, $\adv$, $\cdvi$, $\ldots$) represent new interactions such as those between neutrinos and space-time defects. In particular, QG effects are well-motivated in higher-dimensional operators ($\adv$, $\cdvi$, $\ldots$)~\cite{Weinberg}, whose presence is a sign of an undiscovered high-energy scale, such as the Planck Scale. For example, the Fermi constant associated with a dimension-six operator was one of the first manifestations of electroweak theory. 
These terms correspond to the isotropic part of the Standard-Model Extension (SME)~\cite{Kostelecky:2011gq}, which is an effective field theory that describes the effects of particle Lorentz violation. 
All terms are $3\times 3$ complex matrices in the neutrino flavour basis. 
There are three neutrino flavours: electron-neutrino ($\nue$), muon-neutrino ($\numu$), and tau-neutrino ($\nutau$).
The solution of this Hamiltonian describes the evolution of neutrino flavours.
Because astrophysical flux normalisation is unknown, neutrino flavour is measured in terms of the flavour ratio $(\nue\!:\!\numu\!:\!\nutau)$, which is a normalized fraction of each flavour defined after integrating expected astrophysical neutrino spectra. 
The measured flavour ratio depends on the production mechanisms of astrophysical neutrinos at their sources and on the effective Hamiltonian.
Further details of our formulation are in the Methods. 


The IceCube Neutrino Observatory~\cite{IceCube:2016zyt} is an array of $\NDOM$ digital optical modules (DOMs) embedded in the Antarctic ice between $\BString$~m to $\EString$~m below the surface. 
Each DOM contains one $\PMTsize$~cm photo-multiplier tube in a glass shell, and it detects light from charged particles produced by neutrino interactions.
A series of 60 DOMs are connected with a vertical spacing of $\DOMseparation$~m to make one string, and $\NString$ strings with $\sim\DHole$~m separation cover 1~km$^3$ volume of natural ice as a target volume for astrophysical neutrinos.

When neutrinos undergo charged-current (CC) interactions, they generate charged leptons whose types depend on the neutrino flavours.
Namely, an $\nue$ ($\nuebar$) creates an electron (positron), a $\numu$ ($\numubar$) creates a muon (anti-muon), and a $\nutau$ ($\nutaubar$) creates a tau (anti-tau).
These charged leptons generate characteristic light emission distributions in IceCube.
Electrons initiate electromagnetic showers in ice that look like an approximately isotropic emission of photons (cascade); muons emit light along their straight trajectories (track); and some taus produce an isotropic emission with a slight elongation, reflecting bursts of photon emission from the production of the tau and its subsequent decay (double cascade).
However, most taus from CC interactions and hadronic showers from neutral-current (NC) interactions also lead to cascades.
A likelihood function is constructed from the time and charge distributions of DOMs to estimate the energies, directions, and flavours of neutrinos.
Charged leptons and charged anti-leptons have indistinguishable light emission profiles in ice.


In this analysis, we use the high-energy starting event (HESE) sample with 7.5 years of data collection during 2010 to 2018~\cite{IceCube:2020wum}.
A total of $\Nevt$ events are observed above 60~TeV.
Among them, $\Ncas$, $\Ntrk$, and $\Ndub$ events are classified as cascades, tracks, and double cascades, respectively. 
At most, ten of the lowest energy events are expected to be atmospheric neutrino foreground. 
Cascades and tracks are distributed in $\NbinC$ incoming zenith angle bins, in the range $\cos\th_z=[\LbinC,\HbinC]$, with $\cos\th_z=\HbinC$ pointing to the celestial south pole. We use $\NbinE$ natural logarithmic bins in deposited energy in the range  $E=[\LbinE\,\textrm{TeV},\HbinE\,\textrm{PeV}]$. For the double cascade events, there are $\NbinL$ bins in the reconstructed distance between two cascade signals $L=[\LbinL\,\textrm{m},\HbinL\,\textrm{m}]$ instead of zenith angle bins. 
The Median neutrino energy (zenith angle) resolutions for reconstructed cascades, tracks, and double cascades are 11\%~(6.3$^{\circ}$), 30\%~(1.5$^{\circ}$), and 18\% ~(5.0$^{\circ}$) respectively.


The expected number of events in each bin is computed through a Monte Carlo (MC) simulation.
First, the astrophysical neutrino flux is modeled as a single power-law spectrum.
This is weighted with the assumed flavour ratio at the source and the mixing probability derived from the effective Hamiltonian including new physics operators (Eq.~\ref{eq:hamiltonian}).
The foreground flux due to atmospheric neutrinos from $\pi$ and $K$-decays~\cite{Honda:2006qj}, charm meson decays~\cite{Bhattacharya:2015jpa}, and atmospheric muons~\cite{Heck:1998vt}, is added to simulate the complete flux arriving at the detector.
Neutrino absorption in the Earth is modeled using a standard Earth density profile~\cite{Dziewonski:1981xy}.
Particles produced by neutrino interactions~\cite{Cooper-Sarkar:2011jtt} are computed using specialised MC ~\cite{IceCube:2020tcq} to output photon signals.

\begin{figure}[h!]
\centering
\includegraphics[width=0.9\columnwidth,trim={0 5pt 0 0,},clip]{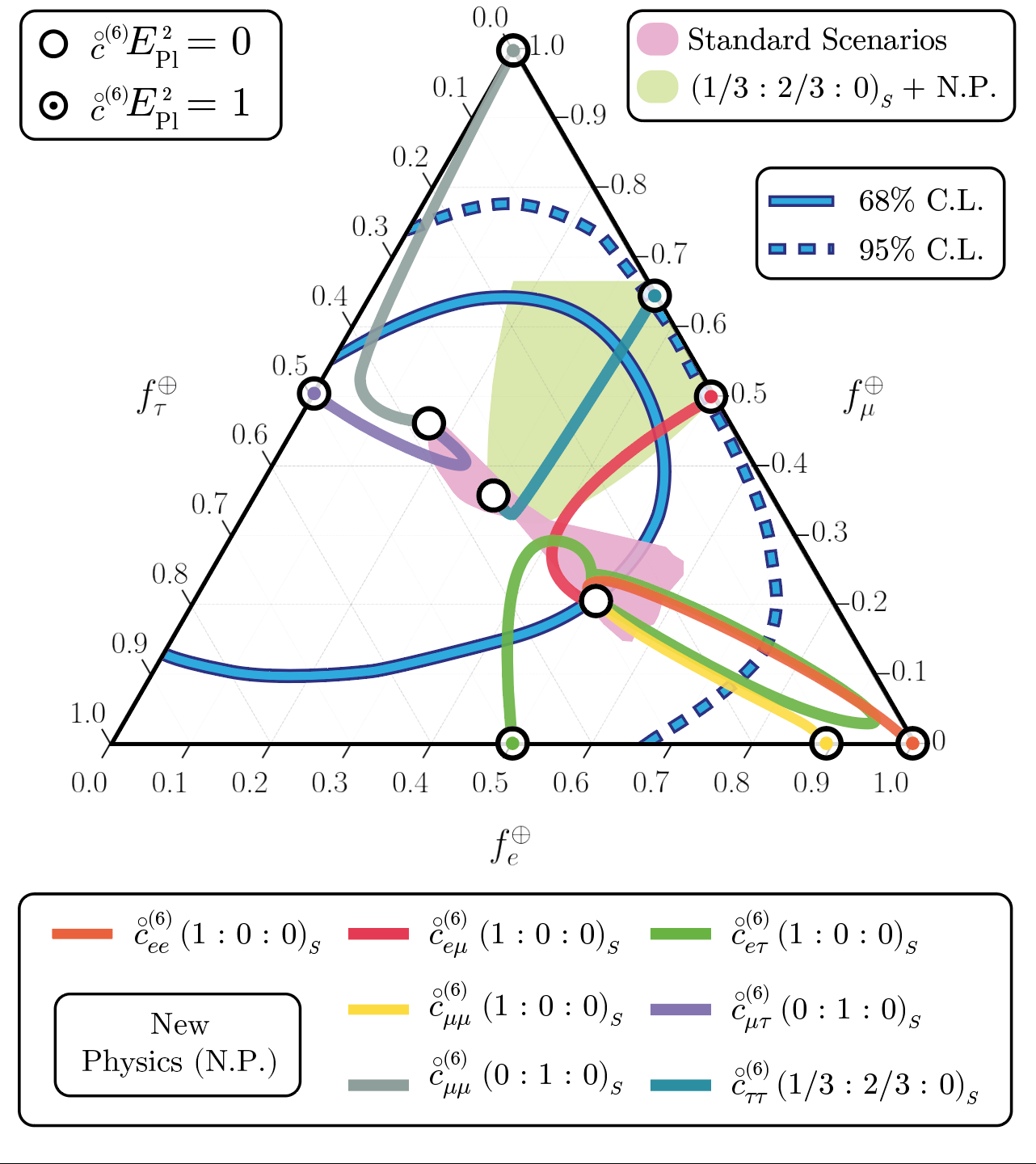}
\caption{
\textbf{\textit{Astrophysical neutrino flavour triangle, including illustrations of new physics effects and data contours.}}
The figure represents the flavour ratio  $(\nue:\numu:\nutau)$ of given compositions at the source ($S$), where the corners indicate pure $\nue$, $\numu$, or $\nutau$ composition. 
The blue solid and dashed lines show $\conone$\% and $\contwo$\% C.L. contours~\cite{IceCube:2020abv} from IceCube data. 
The pink region represents expected flavour ratios from the standard astrophysical neutrino production models, where the neutrinos at the production source are all possible combinations of $\nue$ and $\numu$ with the neutrino oscillation parameter errors given in~\cite{Esteban:2020cvm}. 
The lines explained in the lower legend illustrate the effects of the $\protect\cdvi$ new physics (NP) operators. 
Three astrophysical neutrino production models are highlighted by $\Scale[1.5]{\bigcirc}$ symbols, a $\numu$ dominant source $\muratio$ (top), a $\nue$ dominant source $\eratio$ (bottom), and a preferred model $\piratio$ (middle).  
When NP operators are small ($\le m^2/2E$), they are distributed within the central region.
If the values of NP operators are increased, predicted flavour ratios start to move away from the centre, and they reach to $\Scale[2]{\odot}$ symbols with the large NP such as $\protect\cdvi=E_P^{-2}$. 
For simplicity we concentrate on real, positive new physics potentials.
}
\label{fig:BSM_triangle_0}
\end{figure}


Fig.~\ref{fig:BSM_triangle_0} shows a comparison of the HESE 7.5-yr flavour ratio measurement~\cite{IceCube:2020abv} with model predictions.
This flavour triangle diagram represents astrophysical neutrino flavour ratios where one point in this diagram shows the energy-averaged flavour composition at Earth. 
The pink region near the centre denotes the so-called standard scenarios. 
This represents all possible flavour ratios at Earth from standard astrophysical neutrino production mechanisms via neutrino mixing~\cite{Song:2020nfh}.
As shown, all of the standard flavour ratios are enclosed in the $\contwo$\% confidence level (C.L.) contour, which implies that, at this moment, all models within standard scenarios are allowed.
In other words, the IceCube HESE flavour measurement is consistent with the standard scenarios, given current statistics and systematic errors.
However, current data excludes certain QG models that produce flavour compositions far away from the standard region because any new structure in the vacuum would produce detectable anomalous flavour ratios, shown by lines in Fig.~\ref{fig:BSM_triangle_0}. 

In order to make quantitative statement on these scenarios, we perform a likelihood analysis and report results using a Bayesian method. Our analysis includes all of the flux components previously discussed in the text and implements their systematics according to the prescription given in~\cite{IceCube:2020wum}.
Our analysis likelihood includes: nuisance parameters to incorporate the flux and detector uncertainties, standard oscillation parameters and neutrino mass differences, and parameters that incorporate the QG effective operators.
Technical details of the fit methods and on the systematic errors can be found in the Methods.

\begin{figure}[h!]
\centering
\includegraphics[width=1.0\columnwidth,
trim={0.5cm 0.3cm 0 0},
clip
]{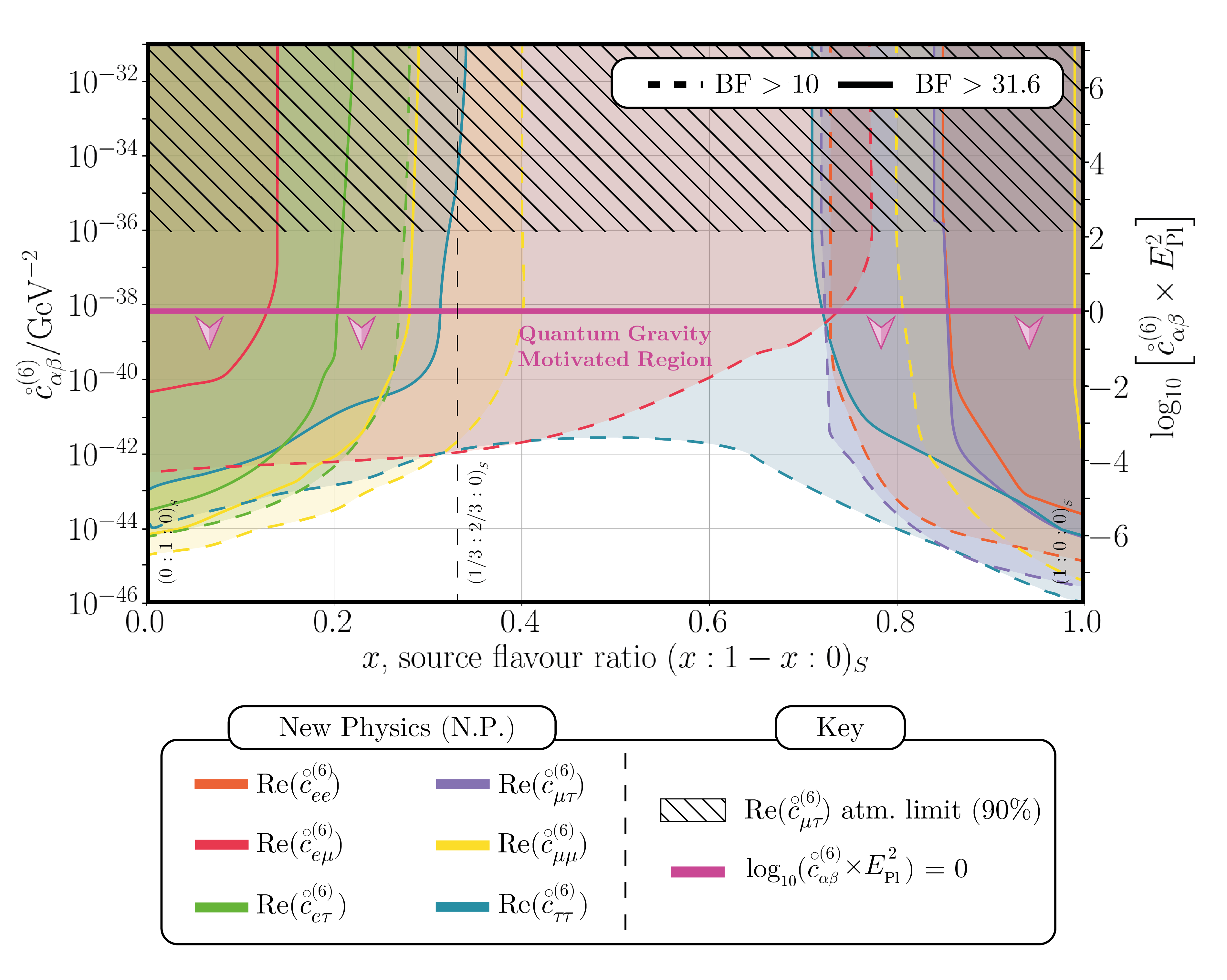}
\caption{
\textbf{\textit{Limits on the dimension-six new physics operator.}}
The QG-motivated physics signal region is defined by $\log_{10}(\protect\cdvi\cdot E_P^2)<0$.
The hatched region is the limit obtained from the atmospheric neutrino data analysis on $\Re(\protect\cutdvi)$~\cite{IceCube:2017qyp}. BF
Limits are presented as a function of the assumed astrophysical neutrino flavour ratio at the production source.
The leftmost scenario is $\numu$ dominant $\muratio$ and the rightmost is $\nue$ dominant $\eratio$.
The preferred scenario corresponds to $\piratio$ (dashed vertical line). 
Limits on $\Re(\protect\ceedvi)$ (orange), 
$\Re(\protect\ceudvi)$ (red), 
$\Re(\protect\cetdvi)$ (green), 
$\Re(\protect\cuudvi)$ (yellow), 
$\Re(\protect\cutdvi)$ (purple), 
and $\Re(\protect\cttdvi)$ (blue) are shown. 
}
\label{fig:final}
\end{figure}

Figure~\ref{fig:final} shows results for the dimension-six operators. 
Results of other operators are summarized in the Supplementary Information.
These represent new physics interactions and we expect the QG-motivated physics operator to be of order $E_P^{-2}=6.7\times 10^{-39}~{\rm GeV}^{-2}$.
Limits are shown on a log-scale. 
The right axis incorporates an additional $E_P^2$ factor where below zero corresponds to the QG-motivated physics signal region. 
For the first time, we reach the QG-motivated signal region of the dimension-six operator with neutrinos. 
The limits are a function of the astrophysical neutrino production model at the source.

Strong limits are obtained for $\numu$ dominant $\muratio$ and $\nue$ dominant $\eratio$ scenarios. 
The $\numu$-dominant scenarios are expected in accelerators such as AGN, while $\nue$-dominant scenarios are expected in accelerators such as neutron stars~\cite{Hummer:2010ai}. 
Substantial limits for $\Re(\protect\cttdvi)$ are obtained across all astrophysical neutrino models including the preferred scenario $\piratio$ which is based on astrophysical pion and muon decays. 
Muon energy loss in the source can cause the $\nue$ fraction to be lower than 1/3 where our limits are valid. The shapes of the limit curves in Fig.~\ref{fig:final} are understood from Fig.~\ref{fig:BSM_triangle_0}. Here, nonzero $\Re(\protect\cetdvi)$ with $\eratio$ and  $\Re(\protect\cutdvi)$ with $\muratio$ models are enclosed by the contour. This suggests that the data cannot set limits on these scenarios. On the other hand, the data reject certain scenarios e.g., a nonzero $\Re(\protect\ceedvi)$ model or a $\Re(\protect\cuudvi)$ model with $\eratio$. 
For the preferred scenario $\piratio$, any new physics scenario is described by the green region, and is almost enclosed by the $\contwo$\% C.L. contour. Given this, only a few substantial limits can be achieved e.g., $\Re(\protect\ceudvi)$ and $\Re(\protect\cttdvi)$. 

In Table~\ref{tab:result}, scenario-independent limits obtained from BF $>\BFSubstantial$ are quoted. These are defined by taking into account a full range of allowed standard astrophysical neutrino production models. Although the motivation of this analysis is to look for evidence of QG, the formalism we have used is model-independent, and our results can set limits on various new physics models~\cite{Rasmussen:2017ert}, including a new long-range force~\cite{Bustamante:2018mzu}, neutrino-dark energy coupling~\cite{Klop:2017dim}, neutrino-dark matter scattering~\cite{Farzan:2018pnk}, violation of equivalent principle~\cite{Fiorillo:2020gsb}, \textit{etc}.

In summary, we have performed the seminal work using astrophysical neutrino flavour information to search for the footprint of QG.
We have not found any evidence of QG, but for the first time, we have reached QG-motivated parameter space for dimension-six operators.
In doing so, we have placed the strongest limits on effective operators that parameterize QG effects across all fields of science.

\def\supbaylimdimiiitattfowo{2 \times 10^{-26}}
\def\supbaylimdimivtcttfowo{2 \times 10^{-31}}
\def\supbaylimdimvtattfowo{2\times 10^{-37}}
\def\supbaylimdimvitcttfowo{3\times 10^{-42}}
\def\supbaylimdimviitattfowo{3\times 10^{-47}}
\def\supbaylimdimviiitcttfowo{2\times 10^{-52}}

\begin{table}
    \begin{tabular}{ccc}
    \hline
    dim & coefficient & limit (BF$>\BFSubstantial$) \\
    \hline \hline
    3 & $\Re(\attdiii)$  & $\supbaylimdimiiitattfowo~\GeV$ \\
    4 & $\Re(\cttdiv)$   & $\supbaylimdimivtcttfowo$ \\
    5 & $\Re(\attdv)$    & $\supbaylimdimvtattfowo~\GeV^{-1}$ \\
    6 & $\Re(\cttdvi)$   & $\supbaylimdimvitcttfowo~\GeV^{-2}$ \\
    7 & $\Re(\attdvii)$  & $\supbaylimdimviitattfowo~\GeV^{-3}$ \\
    8 & $\Re(\cttdviii)$ & $\supbaylimdimviiitcttfowo~\GeV^{-4}$ \\
   \hline
    \end{tabular}
\caption{\textbf{\textit{Limits on new physics operators extracted from this analysis.}}
These limits on new physics operators are derived from BF $>\BFSubstantial$ which corresponds to 1 in $\BFSubstantial$ likelihood ratio for a uniform prior. Limits that depend on assumed production source models are listed in Supplementary Information.}
     \label{tab:result}
\end{table}

\iffalse
\begin{table}
    \begin{adjustbox}{width=\columnwidth,center}
    \footnotesize
    \begin{tabular}{c c p{1.9cm} c c p{1.9cm} c}
    \hline
    dim & coefficient & ~~~~~~limit & dim & coefficient & ~~~~~~limit & $(\ifratio)_S$ \\
    \hline \hline
    
    3 & $\Re(\aeudiii)$   & $\baylimdimiiitaeufowo~\GeV$ & 4 & $\Re(\ceudiv)$ & $\baylimdimivtceufowo$ & $\muratio$ \\
      & $\Re(\aetdiii)$   & $\baylimdimiiitaetfowo~\GeV$ &   & $\Re(\cetdiv)$ & $\baylimdimivtcetfowo$ & $\muratio$ \\
      & $\Re(\auudiii)$   & $\baylimdimiiitauufowo~\GeV$ &   & $\Re(\cuudiv)$ & $\baylimdimivtcuufowo$ & $\muratio$ \\
      & $\Re(\attdiii)$   & $\baylimdimiiitattfowo~\GeV$ &   & $\Re(\cttdiv)$ & $\baylimdimivtcttfowo$ & $\muratio$ \\
      & $\Re(\aeediii)$ & $\baylimdimiiitaeefwoo~\GeV$     &   & $\Re(\ceediv)$ & $\baylimdimivtceefwoo$ & $\eratio $ \\
      & $\Re(\autdiii)$ & $\baylimdimiiitautfwoo~\GeV$     &   & $\Re(\cutdiv)$ & $\baylimdimivtcutfwoo$ & $\eratio $ \\
      & $\Re(\attdiii)$ & $\baylimdimiiitattfwoo~\GeV$     &   & $\Re(\cttdiv)$ & $\baylimdimivtcttfwoo$ & $\eratio $ \\
    \hline
    5 & $\Re(\aeudv)$ & $\baylimdimvtaeufowo~\GeV^{-1}$ & 6 & $\Re(\ceudvi)$ & $\baylimdimvitceufowo~\GeV^{-2}$ & $\muratio$ \\
      & $\Re(\aetdv)$ & $\baylimdimvtaetfowo~\GeV^{-1}$ &   & $\Re(\cetdvi)$ & $\baylimdimvitcetfowo~\GeV^{-2}$ & $\muratio$ \\
      & $\Re(\auudv)$ & $\baylimdimvtauufowo~\GeV^{-1}$ &   & $\Re(\cuudvi)$ & $\baylimdimvitcuufowo~\GeV^{-2}$ & $\muratio$ \\
      & $\Re(\attdv)$ & $\baylimdimvtattfowo~\GeV^{-1}$ &   & $\Re(\cttdvi)$ & $\baylimdimvitcttfowo~\GeV^{-2}$ & $\muratio$ \\
      & $\Re(\attdv)$ & $\baylimdimvtattfwto~\GeV^{-1}$ &   & $\Re(\cttdvi)$ & $\baylimdimvitcttfwto~\GeV^{-2}$ & $\piratio$ \\
      & $\Re(\aeedv)$ & $\baylimdimvtaeefwoo~\GeV^{-1}$ &   & $\Re(\ceedvi)$ & $\baylimdimvitceefwoo~\GeV^{-2}$ & $\eratio $ \\
      & $\Re(\autdv)$ & $\baylimdimvtautfwoo~\GeV^{-1}$ &   & $\Re(\cutdvi)$ & $\baylimdimvitcutfwoo~\GeV^{-2}$ & $\eratio $ \\
      & $\Re(\attdv)$ & $\baylimdimvtattfwoo~\GeV^{-1}$ &   & $\Re(\cttdvi)$ & $\baylimdimvitcttfwoo~\GeV^{-2}$ & $\eratio $ \\
    \hline
    7 & $\Re(\aeudvii)$ & $\baylimdimviitaeufowo~\GeV^{-3}$ & 8 & $\Re(\ceudviii)$ & $\baylimdimviiitceufowo~\GeV^{-4}$ & $\muratio$ \\
      & $\Re(\aetdvii)$ & $\baylimdimviitaetfowo~\GeV^{-3}$ &   & $\Re(\cetdviii)$ & $\baylimdimviiitcetfowo~\GeV^{-4}$ & $\muratio$ \\
      & $\Re(\auudvii)$ & $\baylimdimviitauufowo~\GeV^{-3}$ &   & $\Re(\cuudviii)$ & $\baylimdimviiitcuufowo~\GeV^{-4}$ & $\muratio$ \\
      & $\Re(\attdvii)$ & $\baylimdimviitattfowo~\GeV^{-3}$ &   & $\Re(\cttdviii)$ & $\baylimdimviiitcttfowo~\GeV^{-4}$ & $\muratio$ \\
      & $\Re(\attdvii)$ & $\baylimdimviitattfwto~\GeV^{-3}$ &   & $\Re(\cttdviii)$ & $\baylimdimviiitcttfwto~\GeV^{-4}$ & $\piratio$ \\
      & $\Re(\aeedvii)$ & $\baylimdimviitaeefwoo~\GeV^{-3}$ &   & $\Re(\ceedviii)$ & $\baylimdimviiitceefwoo~\GeV^{-4}$ & $\eratio $ \\
      & $\Re(\autdvii)$ & $\baylimdimviitautfwoo~\GeV^{-3}$ &   & $\Re(\cutdviii)$ & $\baylimdimviiitcutfwoo~\GeV^{-4}$ & $\eratio $ \\
      & $\Re(\attdvii)$ & $\baylimdimviitattfwoo~\GeV^{-3}$ &   & $\Re(\cttdviii)$ & $\baylimdimviiitcttfwoo~\GeV^{-4}$ & $\eratio $ \\
   \hline
    \end{tabular}
    \end{adjustbox}
\caption{
     \textbf{\textit{Limits on new physics operators extracted from this analysis.}}
    These limits on new physics operators are derived from BF $>\BFStrong$ which corresponds to 1 in $\BFStrong$ likelihood ratio for an equal prior. They are for characteristic source flavour ratios; $\eratio$ and $\muratio$. We list only operators where limits are set.}
     \label{tab:result}
\end{table}
\fi

\section*{Methods}

The notation of effective operators follows Ref.~\cite{IceCube:2017qyp}. Explicitly, Eq.~\eqref{eq:hamiltonian} can be written in following way,
\beq
\begin{aligned}
H&\sim
\frac{m^2}{2E}
+\left(\begin{array}{ccc}
\accentset{\circ}{a}^{(3)}_{ee}     & \accentset{\circ}{a}^{(3)}_{e\mu}    & \accentset{\circ}{a}^{(3)}_{\ta e} \\
\accentset{\circ}{a}^{(3)*}_{e\mu}  & \accentset{\circ}{a}^{(3)}_{\mu\mu}  &  \accentset{\circ}{a}^{(3)}_{\mu\ta}\\
\accentset{\circ}{a}^{(3)*}_{\ta e} & \accentset{\circ}{a}^{(3)*}_{\mu\ta} & \accentset{\circ}{a}^{(3)}_{\ta\ta}
\end{array}\right)
-
E\cdot\left(\begin{array}{ccc}
\accentset{\circ}{c}^{(4)}_{ee}     & \accentset{\circ}{c}^{(4)}_{e\mu}    & \accentset{\circ}{c}^{(4)}_{\ta e} \\
\accentset{\circ}{c}^{(4)*}_{e\mu}  & \accentset{\circ}{c}^{(4)}_{\mu\mu}  & \accentset{\circ}{c}^{(4)}_{\mu\ta} \\
\accentset{\circ}{c}^{(4)*}_{\ta e} & \accentset{\circ}{c}^{(4)*}_{\mu\ta} & \accentset{\circ}{c}^{(4)}_{\ta\ta}
\end{array}\right)
\\&+~
\cdots~.
\end{aligned}
\eeq
Beyond the standard terms, there are two groups of new coefficients: the $CPT$-odd terms ($\adiii$, $\adv$, $\advii$, $\ldots$) and the $CPT$-even terms ($\cdiv$, $\cdvi$, $\cdviii$, $\ldots$).
The signs follow the convention of the Standard-Model Extension (SME) given in~\cite{Kostelecky:2011gq}. 
The integers in parentheses represent the dimension $d$ of each operator.
Hence, the units of these operators are GeV$^{4-d}$. 
The dimension-three and dimension-four operators are renormalizable, but all other operators are non-renormalizable.  
All effective operators affecting neutrino flavour conversions have Lorentz indices with temporal, spatial, and mixed components in the Sun-centred celestial equatorial frame (SCCEF)~\cite{Kostelecky:2008ts}.
However, the astrophysical neutrino flux assumed in this analysis is the diffuse flux.
Hence, we assume the incoming neutrino directions are uniform.
This averages out any spatial effects and so this analysis is only sensitive to the isotropic flux component. 
This is reflected in our notation by the circles on top of the operators indicating these operators are spatially isotropic.

In general, two or more operators with different dimensions (such as $\accentset{\circ}{a}^{(3)}_{ee}$ and $\accentset{\circ}{c}^{(4)}_{ee}$) may simultaneously affect the astrophysical neutrino flavour ratio.
However, this is only relevant if the operator scales happen to have similar relative size in the energy region of this analysis; we hypothesise this as an unlikely coincidence and here we do not assume this possibility.
To simplify this analysis, we take only one of the operators to be non-vanishing when reporting our results. 
It is also possible to assume two elements from the same dimensional operator (such as $\accentset{\circ}{a}^{(3)}_{ee}$ and $\accentset{\circ}{a}^{(3)}_{e\mu}$). Since all elements are complex numbers, limits can be set for both real and imaginary parts (such as $\Re(\accentset{\circ}{a}^{(3)}_{ee})$ and $\Im(\accentset{\circ}{a}^{(3)}_{ee})$). 
However, the available data statistics does not allow us to fit two operators with identical energy dimension simultaneously. 
Such assumptions were relaxed in Ref.~\cite{IceCube:2017qyp}. 
The data sample is a mixture of neutrinos and antineutrinos, and we cannot distinguish them on an event-by-event basis. 
In order to assess the impact of the ratio of neutrinos to antineutrinos, we tested the impact of neutrino-only and antineutrino-only fits with a toy Monte Carlo. 
The result of this test yields only marginal changes to the results. 
There, it was found that the complex phase had a small effect on the limits.
Hence, for this analysis we search for one non-negative real element of each operator at a time. 

The solution of the effective Hamiltonian (Eq.~\ref{eq:hamiltonian}) is used to compute the flavour ratio.
We follow the procedure outlined in Ref.~\cite{Arguelles:2015dca}.
First, neutrino flavour eigenstates $\ket{\nu_\al}$ can be described by the superposition of Hamiltonian eigenstates, $\ket{\nu_i}$,
\beq
\ket{\nu_\alpha}= \sum_i V_{\alpha i}(E) \ket{\nu_i}~. 
\eeq
Here, $V_{\al i}(E)$ is a unitary transformation that diagonalizes the effective Hamiltonian (Eq.~\ref{eq:hamiltonian}) to describe the mixing of neutrinos.
Given the large baseline traversed and the energies involved, the resulting neutrino oscillation frequencies are very large and are averaged out by the detector energy resolution.
In this regime, the transition probability of neutrinos can be written only through mixing matrix elements, which are the solution of the effective Hamiltonian. Explicitly, we find
\beq
P_{\nu_{\al} \to \nu_{\be}}(E)&=&
\sum_{i}\left|V_{\al i}(E)\right|^2\left|V_{\be i}(E)\right|^2.
\eeq
The introduction of new interactions in vacuum through the new physics operators $\adiii, \cdiv, \adv, \cdvi$, $\ldots$ is imprinted in the mixing matrix element, $V_{\al i}(E)$, which can be determined through the neutrino mixing.
The observable of interest is the neutrino flux of flavour $\be$ at Earth, $\phi^{\oplus}_\be(E)$, and not the neutrino mixing itself. 
Note that the flux composition at Earth also depends on the initial neutrino flux of flavour $\al$ at the source, $\phi^i_\al(E)$.
Furthermore, the small sample of astrophysical neutrinos restricts us to the use of the energy-averaged flavour composition, 
\beq
\bar \phi^{\oplus}_\be = \frac{1}{|\Delta E|}
\int_{\De E} \sum_\al\bar P_{\nu_{\al}\to\nu_{\be}}(E) \phi^i_\al(E) dE,
\eeq
where we assume a single power-law spectrum for the production flux of astrophysical neutrinos and integrate it.
Finally, we calculate the flavour ratio of astrophysical neutrino flavour $\be$ on Earth by normalizing it, \textit{i.e.},
\beq
 f^{\oplus}_\be=\bar\phi^{\oplus}_\be/\sum_\ga \bar\ph^{\oplus}_\ga~.
\eeq

In this analysis, a total of 14 systematic error nuisance parameters are simultaneously constrained.
Firstly, the following six oscillation parameters~\cite{Esteban:2020cvm}: two neutrino mass-square differences, 
$\De m^2_{21}=\DmSolC^{\DmSolH}_{\DmSolL}$~($\times 10^{-5}$~eV$^2$),  
$\De m^2_{31}=\DmAtmC^{\DmAtmH}_{\DmAtmL}$~($\times 10^{-3}$~eV$^2$); 
three mixing angles, 
$\sin^2 \th_{12}=\ThSolC^{\ThSolH}_{\ThSolL}$,  
$\sin^2 \th_{23}=\ThAtmC^{\ThAtmH}_{\ThAtmL}$, and
$\sin^2 \th_{31}=\ThReaC^{\ThReaH}_{\ThReaL}$; and the 
Dirac $CP$-violating phase, 
$\de_{CP}$ (no constraint). 
Second are five flux systematics which can be classified into two categories: the normalization of each flux component, and the spectral index assuming a single power law.
The normalization systematic errors are introduced as shifts from the nominal predictions, including astrophysical neutrino flux ($\Phi_{\texttt{astro}}$, no constraint), atmospheric neutrino conventional flux ($\Phi_{\texttt{conv}}$, $\SNconvEE$\%), prompt flux ($\Phi_{\texttt{prompt}}$, no constraint), and atmospheric muon flux ($\Phi_{\texttt{muon}}$, $\SNmuonEE$\%). 
The astrophysical neutrino spectral index ($\gamma_{\texttt{astro}}$, no constraint) is also included as a systematic error, where this analysis returns a similar best fit value as in a dedicated study of the same sample~\cite{IceCube:2020wum}. 
We also introduce three detector systematic parameters: the DOM overall efficiency ($\epsilon_{\texttt{DOM}}$, $\SDOMEE$\%), DOM angular dependence ($\epsilon_{\texttt{head-on}}$, $\SDOMangEE$\%), and the in-ice photon propagation anisotropy around DOMs ($a_{\texttt{s}}$, $\SIceEE$\%). 
Additional systematic errors arising from the modelling of atmospheric neutrinos and cosmic rays are used in other analyses~\cite{IceCube:2020wum,IceCube:2020abv}. These systematics are not considered in this analysis because they mostly affect low-energy events ($<$100~TeV). Here, the limit we set for QG-motivated physics depends on the highest end tail of the event distribution.

To place our limits we use two independent analysis methods based on frequentist and Bayesian approaches~\cite{Mandalia:2020lhz,KareemThesis}. 
The Bayesian approach was chosen to be the official result of this analysis due to its accuracy. 
The faster frequentist approach was used to check the Bayesian results.
Both methods use the same 15-dimensional likelihood function with 14 systematic errors where one parameter represents a new physics scale. 
In the Bayesian case, the evidence is obtained by marginalising the same likelihood over the systematic parameters via a nested sampling.
The marginalisation is done assuming model priors from ~\cite{IceCube:2020wum}. 
We use the MultiNest algorithm~\cite{Feroz:2008xx} with $\livep$ live points, approximately $\chain$ steps, and a tolerance of $\toler$. 
An example of a posterior distribution from one configuration is shown in the Supplementary Information. Here, the log-likelihood is calculated for $\Re(\protect\accentset{\circ}{c}^{(6)}_{\tau\tau})=10^{-44}~\GeV^{-2}$, with an assumed source flavour ratio $\eratio $. We run nearly $\sim 200,000$ similar calculations with different configurations to map out the parameter phase space to find the signal. 

We then define the BF to be the ratio of the model evidence with respect to the null hypothesis. We use Jeffreys' scale to set substantial and strong limits that are defined by the BF to be larger than $\BFSubstantial$ (substantial limit) and $\BFStrong$ (strong limit). 
Supplementary Information includes 
an example of such a plot. Here, multiple sample runs are combined to construct a BF distribution with a function of $\Re(\protect\accentset{\circ}{c}^{(6)}_{\tau\tau})$ with an assumed source flavour ratio $\eratio$. Due to limited statistics, simulated points have errors, and we use a spline function to extrapolate values between points.  Supplementary Information list the substantial and strong limits from this procedure for selected source flavour ratio assumptions. Because of the extrapolation we used, we do not have enough accuracy to set limits and our results are presented with one digit. 

We repeat a similar BF function construction with different source flavour ratio assumptions, and we construct the limits with a function of source flavour ratio $x$ (Fig.~\ref{fig:final}). We use a spline function to extrapolate values between points and smooth the limit lines.  
We repeat this for different dimension operators, and these results are shown in 
the Supplementary Information. The limits of this analysis go stronger for higher dimensions due to the stronger energy dependence (Eq.~\eqref{eq:hamiltonian}). As the dimension increases, limits from astrophysical neutrino interferometry become significantly stronger than atmospheric neutrino interferometry. On the other hand, for dimension seven and eight operators, QG-motivated physics is expected to be smaller than $E_P^{-3}$ and $E_P^{-4}$. This analysis has no sensitivity to the operators with these sizes. The dimension-three and -four operators are renormalizable and we cannot define QG-motivated physics in the same way.
In the frequentist case, the likelihood function is used to find the best-fit point. Then, the profile likelihood ratio of the best fit to the null hypothesis is used to set limits assuming Wilks' theorem. 
However, Wilks' theorem may not hold in the full likelihood space which affect the limits obtained. 
We use distributed Open Science Grid~\cite{Pordes:2007zzb} computational resources to perform both analyses.

\section*{Data availability}
The data events and simulation used in this analysis are described in Ref.~\cite{IceCube:2020wum}, and files are available
at https://github.com/icecube/HESE-7-year-data-release.

\section*{Code availability}
Much of the analysis code is IceCube proprietary and exists as part of the
IceCube simulation and production framework. IceCube open-source
code can be found at https://github.com/icecube. Additional information
is available from analysis@icecube.wisc.edu upon request.

\section*{Acknowledgements}
The authors gratefully acknowledge the support from the following agencies and institutions: 
USA {\textendash} U.S. National Science Foundation-Office of Polar Programs,
U.S. National Science Foundation-Physics Division,
U.S. National Science Foundation-EPSCoR,
Wisconsin Alumni Research Foundation,
Center for High Throughput Computing (CHTC) at the University of Wisconsin{\textendash}Madison,
Open Science Grid (OSG),
Extreme Science and Engineering Discovery Environment (XSEDE),
Frontera computing project at the Texas Advanced Computing Center,
U.S. Department of Energy-National Energy Research Scientific Computing Center,
Particle astrophysics research computing center at the University of Maryland,
Institute for Cyber-Enabled Research at Michigan State University,
and Astroparticle physics computational facility at Marquette University;
Belgium {\textendash} Funds for Scientific Research (FRS-FNRS and FWO),
FWO Odysseus and Big Science programmes,
and Belgian Federal Science Policy Office (Belspo);
Germany {\textendash} Bundesministerium f{\"u}r Bildung und Forschung (BMBF),
Deutsche Forschungsgemeinschaft (DFG),
Helmholtz Alliance for Astroparticle Physics (HAP),
Initiative and Networking Fund of the Helmholtz Association,
Deutsches Elektronen Synchrotron (DESY),
and High Performance Computing cluster of the RWTH Aachen;
Sweden {\textendash} Swedish Research Council,
Swedish Polar Research Secretariat,
Swedish National Infrastructure for Computing (SNIC),
and Knut and Alice Wallenberg Foundation;
Australia {\textendash} Australian Research Council;
Canada {\textendash} Natural Sciences and Engineering Research Council of Canada,
Calcul Qu{\'e}bec, Compute Ontario, Canada Foundation for Innovation, WestGrid, and Compute Canada;
Denmark {\textendash} Villum Fonden and Carlsberg Foundation;
New Zealand {\textendash} Marsden Fund;
Japan {\textendash} Japan Society for Promotion of Science (JSPS)
and Institute for Global Prominent Research (IGPR) of Chiba University;
Korea {\textendash} National Research Foundation of Korea (NRF);
Switzerland {\textendash} Swiss National Science Foundation (SNSF);
United Kingdom {\textendash} Department of Physics, University of Oxford, the Royal Society, and the Science and Technology Facilities council (STFC).

\section*{Author contributions}
The IceCube collaboration acknowledges the significant contributions to this manuscript from Carlos Arg\"{u}elles, Kareem Farrag, and Teppei Katori. The IceCube Collaboration designed, constructed and now operates the IceCube Neutrino Observatory. Data processing and calibration, Monte Carlo simulations of the detector and of theoretical models, and data analyses were performed by a large number of collaboration members, who also discussed and approved the scientific results presented here.


\end{document}